\documentclass{article}
\usepackage{setspace}
\title{NEUTRINOS AS SOURCE OF ULTRA HIGH ENERGY COSMIC RAYS IN EXTRA DIMENSIONS}

\author{Ashok Goyal\thanks{E--mail : agoyal@ducos.ernet.in},
Abhinav Gupta\thanks{E--mail : abh@ducos.ernet.in, abhinav@physics.du.ac.in},
Namit Mahajan\thanks{E--mail : nm@ducos.ernet.in, nmahajan@physics.du.ac.in}\\
	{\em Department of Physics and Astrophysics,} \\
	 {\em University of Delhi, Delhi-110 007, India.}}

\begin{document}
\doublespacing
\maketitle
\large
\newcommand{\del}{\mbox{$\partial$}} 
\begin{abstract}

If the neutrinos are to be identified with the primary source of ultra-high energy cosmic rays(UHECR), their interaction on relic neutrinos is of great importance in understanding their long intergalactic journey. In theories with large compact dimensions, the exchange of a tower of massive spin-2 gravitons (Kaluza-Klein excitations) gives extra contribution to $\nu\bar{\nu} \longrightarrow f\bar{f}$ and $\gamma\gamma$ processes along with the opening of a new channel for the neutrinos to annihilate with the relic cosmic neutrino background $\nu\bar{\nu} \longrightarrow G_{kk}$ to produce bulk gravitons in the extra dimensions. This will affect their attenuation. We compute the contribution of these Kaluza-Klein excitations to the above processes and find that for parameters of the theory constrained by supernova cooling, the contribution does indeed become the dominant contribution above $\sqrt{s}~\simeq~ 300$ GeV.
\end{abstract} 
\pagebreak

Ultra high energy (UHE) cosmic rays \cite{Takeda} present a challange in cosmology and astroparticle physics. Energies beyond the Greisen-Zatsepin-Kuzmin (GZK) cut off $\sim 5\times10^{19}$ eV excludes protons as primary candidates \cite{Greisen} for high energy events due to the fact that above the resonance threshold for $\Delta^{\ast}$ production, the protons would lose roughly 20 $\%$ of their energy by scattering on the $2.73~K$ cosmic background photons through $p + {\gamma_{2.73K}}\longrightarrow{\Delta^{\ast}}\longrightarrow N + \pi$ during 6 Mpc (mean free path) of their journey. Thus apart from the difficulty in accelerating particles to such extremely high energies \cite{Biermann}, these protons would lose significant fraction of their energy if they came from 100 Mpc or further away. Gamma rays and neutrinos could be other primary candidates which will not be hampered by the GZK cut off. The high energy photon annihilation cross-section on cosmic background photons \cite{Berezinskii} will however, preclude them from being the candidates particularly if they have to travel a distance of the order of 100 Mpc of the intergalactic space, their mean free path being of the order of 10-40 Mpc only. Weiler \cite{Weiler} proposed that if neutrinos have to be the source of UHE cosmic rays, they have to have their energies corresponding to the Z-resonance in order to be able to effeciently annihilate relic neutrinos i.e. $E_{\nu}\simeq\frac{{M_Z}^2}{2M_{\nu}}~\simeq~ 10^{23}$ eV for neutrino mass consistent with the Super-Kamiokanda data which is clearly above the GZK cut off \cite{Gelmini}. The difficulty in realizing this scenario is to identify the source of ultra high energy neutrinos with their energy close to the Z-resonance. Decay of a super heavy relic particle into neutrinos with mass $M_{\chi} ~\geq~ 10^{13}$ GeV has been proposed \cite{Gelmini} to be the source of these highly energetic neutrinos that can explain cosmic ray events beyond the GZK cut off. Such neutrinos however, have to be produced in a spherical shell at the red shift $z = (\frac{M_{\chi}}{2E_{res}} -1) = (\frac{{M_{\chi}}{M_{\nu}}}{{M_Z}^2} -1)$ so that their energy near earth is close to the Z-resonance energy for the $\nu\bar{\nu}$ annihilation cross-section on the relic neutrinos to be large.If such fine tuning of mass of the relic super heavy particle or their abundance at the red shift $z = (\frac{M_{\chi}}{2E_{res}} -1)$ is not available, we will require large neutrino hadron cross-sections of hadronic strength for the neutrinos to be able to initiate showers high in the atmosphere. Current estimates of the ultra high neutrino scattering in the Standard Model for $E_{\nu} ~\sim~ 10^{21}$ eV are estimated \cite{Gandhi} to lie in the range $10^{-4}~-~10^{-5}$ mb. Specifically,

\begin{equation}
{\sigma_{{\nu}N}}^{SM}(E_{\nu}) \simeq (2~-~3){\times 10^{-5}} (\frac{E_{\nu}}{10^{19}eV})^{0.4}  ~mb
\end{equation}

The anomalously large high energy neutrino interaction requires new physics and an appealing possibility considered in the literature is the theory with n extra dimensions with large compactification radii and TeV scale quantum gravity \cite{Arkani}. In such theories, exchange of a tower of massive spin-2 bulk gravitons (Kaluza-Klein(KK) excitations) opens up the exciting possibility of extra contribution to any two particle scattering. In this context, the contribution of gravitational scattering to ${\nu}N$ cross-section consistent with the requirement of unitarity has been estimated \cite{Nussinov}. For example, for a neutrino of energy $E_\nu$ scattering on a proton at rest

\begin{equation}          
{\sigma_{\nu N}}^{g}(E_{\nu}) \simeq \frac{4{\pi}s}{{M_s}^4} \simeq 0.1 (\frac{1TeV}{M_S})^4 (\frac{E_{\nu}}{10^{19}eV}) ~ mb
\end{equation}

The most stringent lower limits on the string scale $M_S$ are obtained from astrophysical considerations \cite{Callen} by studying the energy loss rates through the emission of bulk gravitons in SN 1987 A and lead to typically $M_{S} \geq$ 30 Tev, 4 TeV and 1 TeV for the number of extra dimensions equal to 2, 3 and 4 respectively. Thus n~=~4 is required if neutrinos are to be the primary candidates for UHE cosmic ray events.\\
The new physics would open up new channels for these high energy neutrinos to annihilate with the relic neutrino background to produce bulk gravitons in the extra dimensions through
\begin{equation}
\nu\bar{\nu} \longrightarrow G_{KK}
\end{equation}
and may therefore get effeciently attenuated during their intergalactic journey. In these theories, there also exists extra contributions to high energy neutrino scattering on relic background neutrinos through the exchange of KK excitations to produce the cascade of high energy particles through  

\begin{equation}
\nu\bar{\nu} \longrightarrow f\bar{f}
\end{equation}
and
\begin{equation}
\nu\bar{\nu} \longrightarrow \gamma\gamma
\end{equation}
 processes. In the Standard Model the process (4) takes place through the exchange of Z, while the process (5) can take place only through the loop diagrams and is expected to be highly suppressed \cite{Gell}. In theories with extra dimensions, there is direct coupling between Standard Model particles and spin-2 gravitons, thus providing a unique channel.\\
Using the Feynman Rules given in  \cite{Han}, it is straight forward to calculate the above processes. The spin averaged neutrino annihilation cross-section to produce a KK graviton can be easily found to be
\begin{equation}
\sigma(\nu \bar{\nu} \longrightarrow G_{KK}(s,m_n)) = \frac{\pi~\kappa^2~\sqrt{s}}{32}~\delta~(m_n~-~\sqrt{s})
\end{equation}
where $m_n$ is the mass of the nth KK state, being given by $m_n^2 = \frac{4\pi^2n^2}{R^2}$ where $\kappa^2~R^n = 16\pi~{(4\pi)}^{\frac{n}{2}}~\Gamma(\frac{n}{2})~M_s^{-(n+2)}$. Summing over all the KK states using the density of the number of states given in equations (B2) and (B3) in \cite{Han}, the total cross-section comes out to be
\begin{equation}
\sigma(\nu \bar{\nu} \longrightarrow G_{KK}) = \frac{\pi^{2}}{s}
(\frac{s}{M_{s}^{2}})^{{\frac{n}{2}}+1} 
\end{equation}
where n is the number of extra dimensions.\\
The matrix elements for the scattering processes (4) and (5) are given in \cite{Han} (see equations (61), (62), (71) and (72)). Squaring the matrix elements and summing over the virtual KK state exchange propagator using equations (B4)-(B8) of \cite{Han}, the total scattering cross-section can be calculated in a straight forward manner and we get
\begin{equation}
\sigma^{g}(\nu\bar{\nu} \longrightarrow f\bar{f}) = \frac{\pi}{60s} {(\frac{s}{{M_s}^2})}^{n+2} {\cal{F}}^2
\end{equation}
and 
\begin{equation}
\sigma^{g}(\nu\bar{\nu} \longrightarrow \gamma\gamma) = \frac{\pi}{20s} {(\frac{s}{{M_s}^2})}^{n+2} {\cal{F}}^2
\end{equation}
where ${\cal{F}}^{2} = {\pi}^{2} + 4I^{2}(\frac{{M_s}}{\sqrt{s}})$ and $I$ is given in \cite{Han} (see equation (B8)).\\
In these calculations it was assumed that the massless four dimensional graviton and its massive KK excitations couple with the same strength and the sum over all KK states to a given scattering amplitude in general is obtained by introducing a cut-off. However, it has recently been pointed out \cite{Bando} that due to brane fluctuations, effective coupling of the nth KK mode to four dimensional fields is suppressed exponentially and is given as
\begin{equation}
\kappa_n = \kappa~ \exp\left(\frac{-c~m_n^2}{M_s^2}\right)
\end{equation}
where c parametrises the effects of finite brane tension and is of order one or larger. This suppression then provides a dynamical cut-off in the sum over KK states. With this modified coupling, neutrino-nucleon cross-section due to the exchange of KK gravitons has been recalculated \cite{KachelrieB} and has been found to be considerably smaller to account for UHE cosmic ray observations. If that is so, neutrinos remain deeply penetrating and can be the source of UHE cosmic rays only if their energies correspond to the Z resonance \cite{Weiler}, \cite{Gelmini}.\\
For a neutrino of energy $E_{\nu}$ annihilating a relic neutrino of mass $m_{\nu}$, we get
\begin{equation}
\sigma^{g}(\nu\bar{\nu} \longrightarrow G_{KK}) \simeq
4\times10^{-33-\frac{3n}{2}} 2^{\frac{n}{2}}
\left(\frac{{m_{\nu}}}{1eV}\right)^{\frac{n}{2}} \left(\frac{E_\nu}{10^{21}eV}\right)^{\frac{n}{2}} \left(\frac{1 TeV}{{M_s}}\right)^{n+2} ~ cm^{2}
\end{equation}
and
\begin{equation}
\sigma^{g}(\nu\bar{\nu} \longrightarrow f\bar{f}) \simeq 4\times10^{-38-3n} 2^{n} \left(\frac{m_{\nu}}{1eV}\right)^{n+1} \left(\frac{E_\nu}{10^{21}eV}\right)^{n+1}\left(\frac{1 TeV}{M_s}\right)^{2n+4} ~{\cal{F}}^{2} ~ cm^{2}
\end{equation}
and 
\begin{equation}
{{\sigma}^{g}}(\nu\bar{\nu} \longrightarrow \gamma\gamma) = 3 {{\sigma}^{g}}(\nu\bar{\nu} \longrightarrow f\bar{f})
\end{equation}

to be compared with the Standard Model calculation
\begin{equation}
\sigma^{SM}(\nu\bar{\nu} \longrightarrow f\bar{f}) = \frac{G_F^{2}}{3\pi}\frac{M_Z^{2} s}{{(s-M_Z^{2})}^{2} + M_Z^{2}\Gamma_z^{2}} (C_V^{2}+C_A^{2})
\end{equation}
where ${C_V}^{f} = {T_3}^{f} - 2Q^{f}sin^{2}{{\theta}_{w}}$ and ${C_V}^{f} = {T_3}^{f}$. The Standard Model cross-section at $s = M_Z^{2}$ comes out to be $~\simeq~ 1.59{\times}10^{-32}~ cm^{2}$. The process $\nu\bar{\nu} \longrightarrow \gamma\gamma$ has been recently calculated in the Standard Model \cite{Abba} and is given by
\begin{equation}
\sigma^{SM}(\nu\bar{\nu} \longrightarrow \gamma\gamma) =  \frac{G_F^{2}\alpha^{2}}{640\pi^{3}}\frac{s^{3}}{M_W^{4}} A^{2}\\ \simeq 5.6\times10^{-42}\left(\frac{m_\nu}{1eV}\right)^{3}\left(\frac{E_\nu}{10^{21}eV}\right)^{3} cm^{2}
\end{equation}
where A is taken to be ${\simeq} 14.4$. The above expression is valid for $\sqrt{s}$ upto roughly $2M_W$ after which the cross-section reaches a plateau till about 1 TeV and then starts falling \cite{Abba}.\\
The effect of brane fluctuations through modification of the coupling can be easily incorporated in the calculations. $\sigma(\nu\bar{\nu} \longrightarrow G_KK)$ gets multiplied by a suppression factor $\exp(\frac{-2cs}{M_s^2})$ and in the scattering cross-sections (8) and (9), ${\mathcal{F}}^2$ is replaced by $|{\mathcal{D}}|^2$ where
\begin{equation}
\mathcal{D} = \left[-\iota\pi~\exp\left(-\frac{2cs}{M_s^2}\right) - J\left(\frac{2cs}{M_s^2}\right)\right]
\end{equation}
with
\begin{equation}
J\left(\frac{2cs}{M_s^2}\right) = {\mathcal{P}}\int_0^\infty dy \frac{y^{\frac{n}{2}-1}}{1~-~y}~\exp\left(\frac{-2cy}{M_s^2}\right)
\end{equation}
where $\mathcal{P}$ denotes the Principal Value of the integral and for even values of n it reduces to 
\begin{equation}
J\left(\frac{2cs}{M_s^2}\right) = (-1)^{n-1}~Ei\left(\frac{2cs}{M_s^2}\right) + \sum_{k=1}^{\frac{n}{2}-1}~(k-1)!~\exp\left(\frac{2cs}{M_s^2}\right)~\left(\frac{M_s^2}{2cs}\right)^k
\end{equation}
where $Ei(x)$ is the Exponential-integral function \cite{Grad}.\\
The suppression in the cross-sections due to brane fluctuations at the available values of $\sqrt{s}$ even for the smallest allowed value of $M_s = 1$ TeV for n~=~4 is not more than $10~\%$ and is not significant as can be seen from the expression of $\mathcal{D}$.\\
In Figures 1-3, we have plotted the cross-sections for the cases n = 2 and 4 with $M_s$ = 1, 10 and 30 TeV. In Fig.4, we plot the cross-sectons for the parameters of the theory that are constrained from astrophysical considerations obtained by demanding that the supernova cooling through graviton production does not exceed $10^{53}$ ergs. We have also plotted the Standard Model cross-sections for $\nu\bar{\nu} \longrightarrow f\bar{f}$ and $\nu\bar{\nu} \longrightarrow \gamma\gamma$ for comparison. We see that the contribution of spin-2 graviton exchange for n=2 and $M_s$ = 1 TeV rises very fast with energy and far exceeds the Standard Model contribution. Such a large cross-section would result in the attenuation of high energy neutrinos if they are produced through the decayof a super heavy relic particle of mass $M_\chi$ exceeding $10^{14}$ GeV. The probability of these high energy neutrinos to interact depends upon the cross-section and on the lepton-asymmetry parameter $\eta = \frac{\eta_{\nu,relic}}{\eta_{\gamma,0}}$ whose value is constrained by nucleosynthesis and by large scale structure considerations. The mean free path of neutrinos is dominated by $\sigma(\nu\bar{\nu} \longrightarrow G_{KK})$ and is given by $\lambda = \frac{1}{\sigma(\nu\bar{\nu} \longrightarrow G_{KK}){\eta_{\nu,relic}}}$. If there is no lepton-asymmetry, ${\eta} ~\simeq~ 0.14$ and could be as large as 4 for some cosmological parameters \cite{Adams}. For $\sqrt{s} = 500$ GeV,the neutrino mean free path for n~=~2 and $M_{s} =1$ TeV is $\lambda_{\nu\bar{\nu} \rightarrow G_{KK}}~ \simeq~ 0.57{\times}10^{7} (\frac{0.14}{\eta})$ Mpc to be compared with the corresponding contribution in the Standard Model from the Z-exchange namely $\lambda_{\nu\bar{\nu} \rightarrow Z^{\ast}} ~\simeq~ 1.33{\times}10^{10} (\frac{0.14}\eta)$ Mpc. For the astrophysically prefered value (n~=~4, $M_{s}~=~1$ TeV), the mean free path is $\lambda_{\nu\bar{\nu} \rightarrow G_{KK}} ~\simeq~ 2.27\times10^{7} (\frac{0.14}{\eta})$ Mpc.\\
At resonance, $\sqrt{s} = M_Z$, the Z-exchange gives by far the largest cross-section with mean free path $\lambda_{{\nu\bar{\nu} \rightarrow Z^{\ast}}_{res}} ~\simeq~ 0.345{\times}10^{6} (\frac{0.14}{\eta})$ Mpc to be compared with\\ $\lambda_{\nu\bar{\nu} \rightarrow G_{KK}} ~\simeq~ 6.8{\times}10^{8} (\frac{0.14}{\eta})$ Mpc for the astrophysically allowed parameters.\\
However, neutrinos produced by the decay of super heavy relic particles with $\sqrt{s}~ {\geq}~ 500$ GeV would mainly annihilate to produce bulk gravitons with a cross-section more than three orders of magnitude greater then the Standard Model cross-section through $Z^{\ast}$ exchange. It should however be noted that these new neutrino interactions with CMB photons and massless relic neutrinos would not have enough center of mass energy to compete with the Standard Model interactions to significantly influence the neutrino propagation. For massive neutrinos on the other hand, the relic neutrino background provides large center of mass energy for these interactions to play an important role in neutrino propagation through their intergalactic journey \cite{Tyler}. For example, from Fig.4, we observe that for Super-Kamiokande motivated neutrino mass $~\simeq~ 0.07$ eV, $\nu\bar{\nu}$ annihilation cross-section for bulk graviton production for neutrinos of energy $~\simeq~ 10^{15}$ GeV, ($\sqrt{s} ~\simeq~ 265$ GeV) is comparable to the Standard Model cross-section. It is thus clear that in a scenario in which the neutrinos are produced through the decay of super-heavy relic particles of mass $M_X > 10^{13}$ GeV, the $\nu\bar{\nu}$ annihilation cross-section on the relic neutrinos in the Standard Model is large enough at $\sqrt{s}$ near the Z-resonance energy to qualify for the source of UHE cosmic rays beyond the GZK cut-off even for Super-Kamiokande constrained neutrino mass range. At these energies the new interactions remain much smaller than the Standard Model ones. This of course, requires that the X particles have life time long compared to the age of the universe, $\tau_X \geq 10^5~t_o$. On the other hand, if neutrinos were produced at large z i.e. at large cosmological distances, which would be the case for super-heavy particles with life time much smaller than the age of the universe, large center of mass energy (enhanced by a factor of (1+z)) will be available to the neutrinos for new interactions to play an important and for values of $\sqrt{s} > 265$ GeV even dominant role. The cascade of charged particles and $\gamma$ rays produced through the exchange of a tower of massive KK excitations in theories with n extra dimensions with large compactification scale will give significant contribution to the GeV gamma ray background today.

\begin{section}*{Acknowledgements}
One of us, A.~Goyal, would like to thank the organisers of WHEPP-6,
held in the Institute of Mathematical Sciences, Chennai, India from
Jan. 3-15, 2000 where the problem was discussed in the working
group. A.~G thanks CSIR, India while N.~M. thanks the University Grants Commission, India, for financial support.
\end{section}
\pagebreak

\begin{figure}[ht]
\vskip 15truecm
\includegraphics{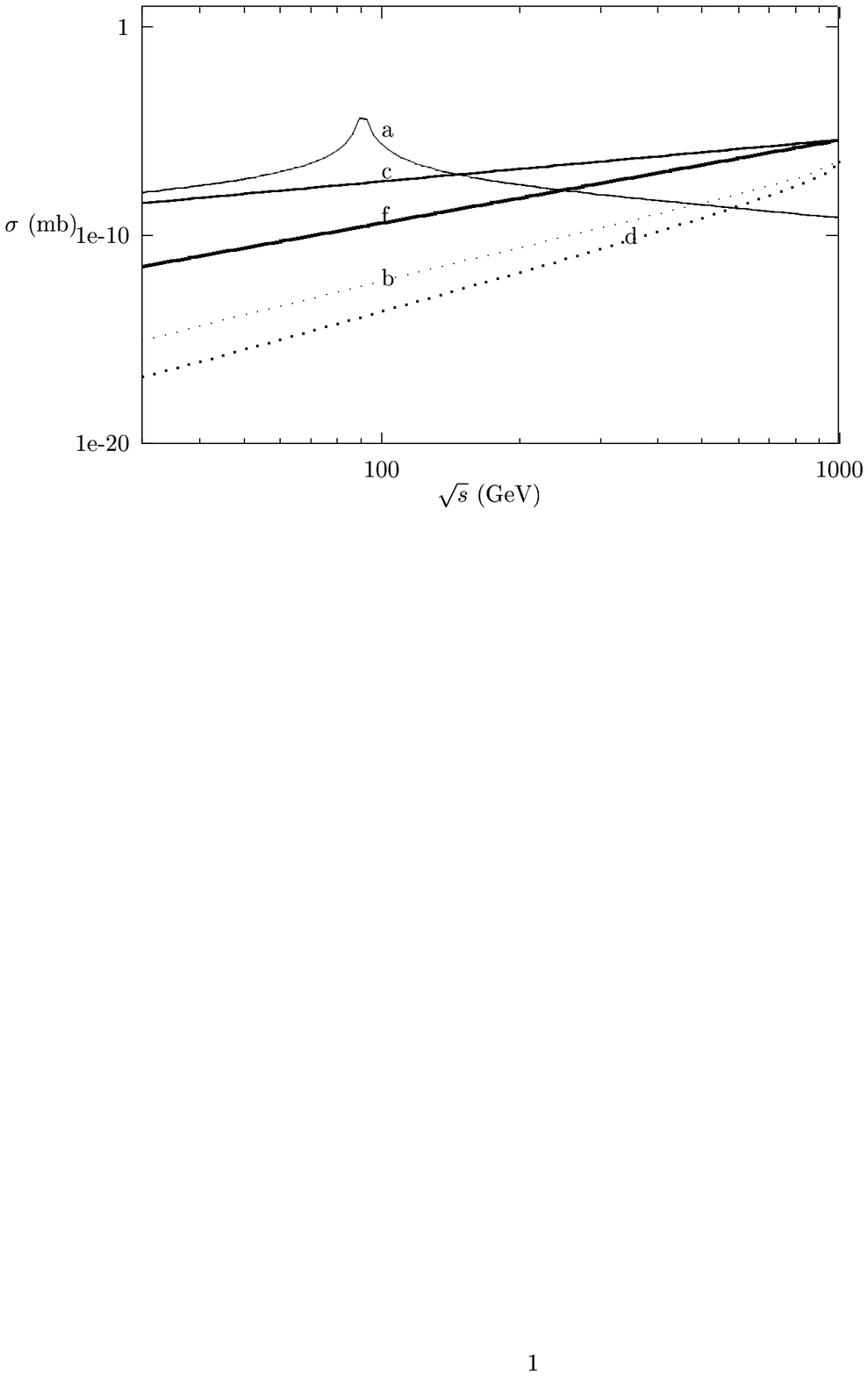}
\caption{Total $\nu\bar{\nu}$ annihilation cross-section as a function of $\sqrt{s}$. Curves a, b, c, d and f stand for $\sigma^{SM}(\nu\bar{\nu} \longrightarrow f\bar{f})$, $\sigma^{g}(\nu\bar{\nu} \longrightarrow f\bar{f}, n=2)$, $\sigma(\nu \bar{\nu} \longrightarrow G_{KK}, n=2)$, $\sigma^{g}(\nu\bar{\nu} \longrightarrow f\bar{f}, n=4)$, $\sigma(\nu \bar{\nu} \longrightarrow G_{KK}, n=4)$ respectively for $M_s$ = 1 TeV.}
\end{figure}

\begin{figure}[ht]
\vskip 15truecm
\includegraphics{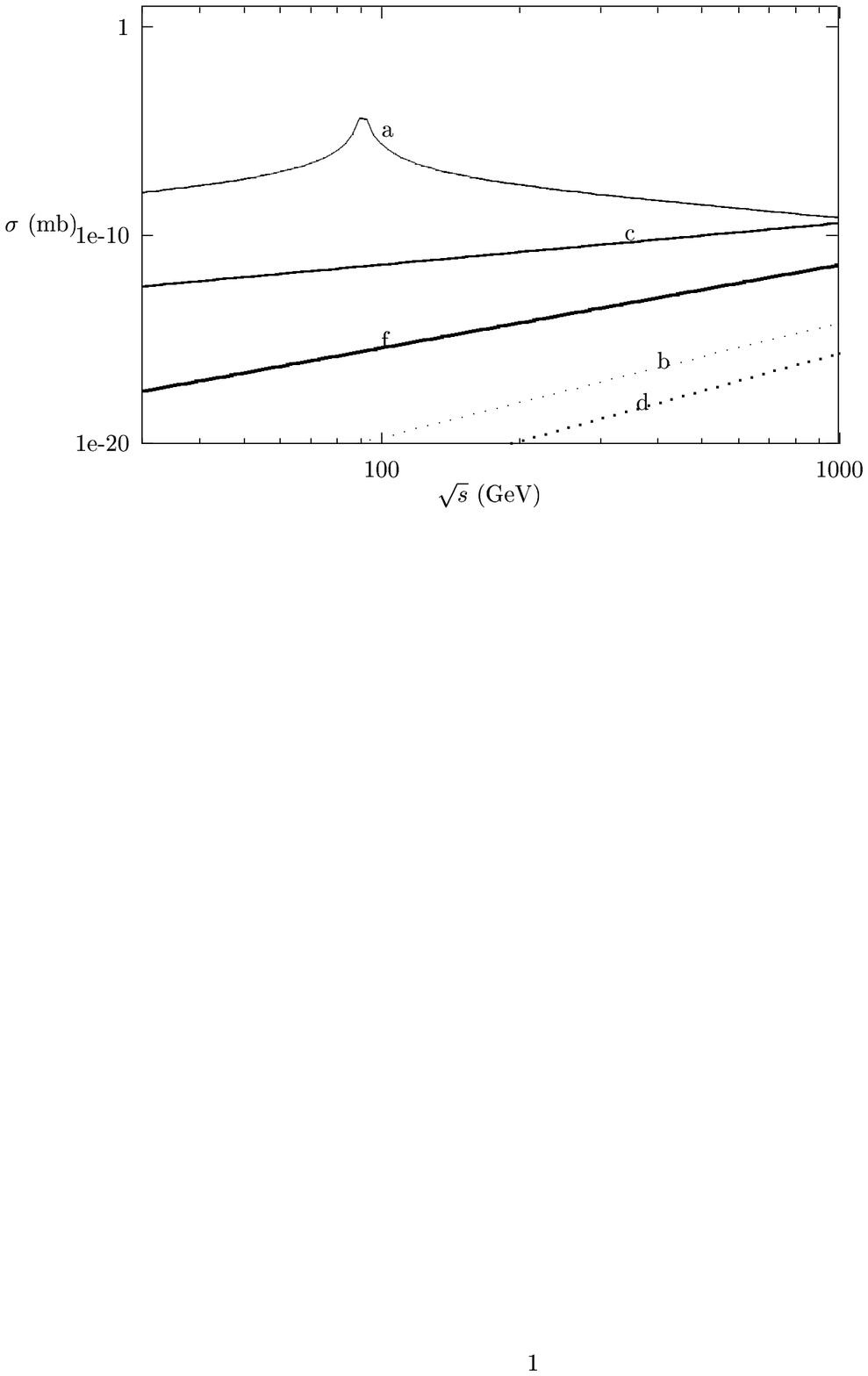}
\caption{Total $\nu\bar{\nu}$ annihilation cross-section as a function of $\sqrt{s}$. The curves are as in Fig.1 and $M_s$ = 10 TeV.}
\end{figure}

\begin{figure}[ht]
\vskip 15truecm
\includegraphics{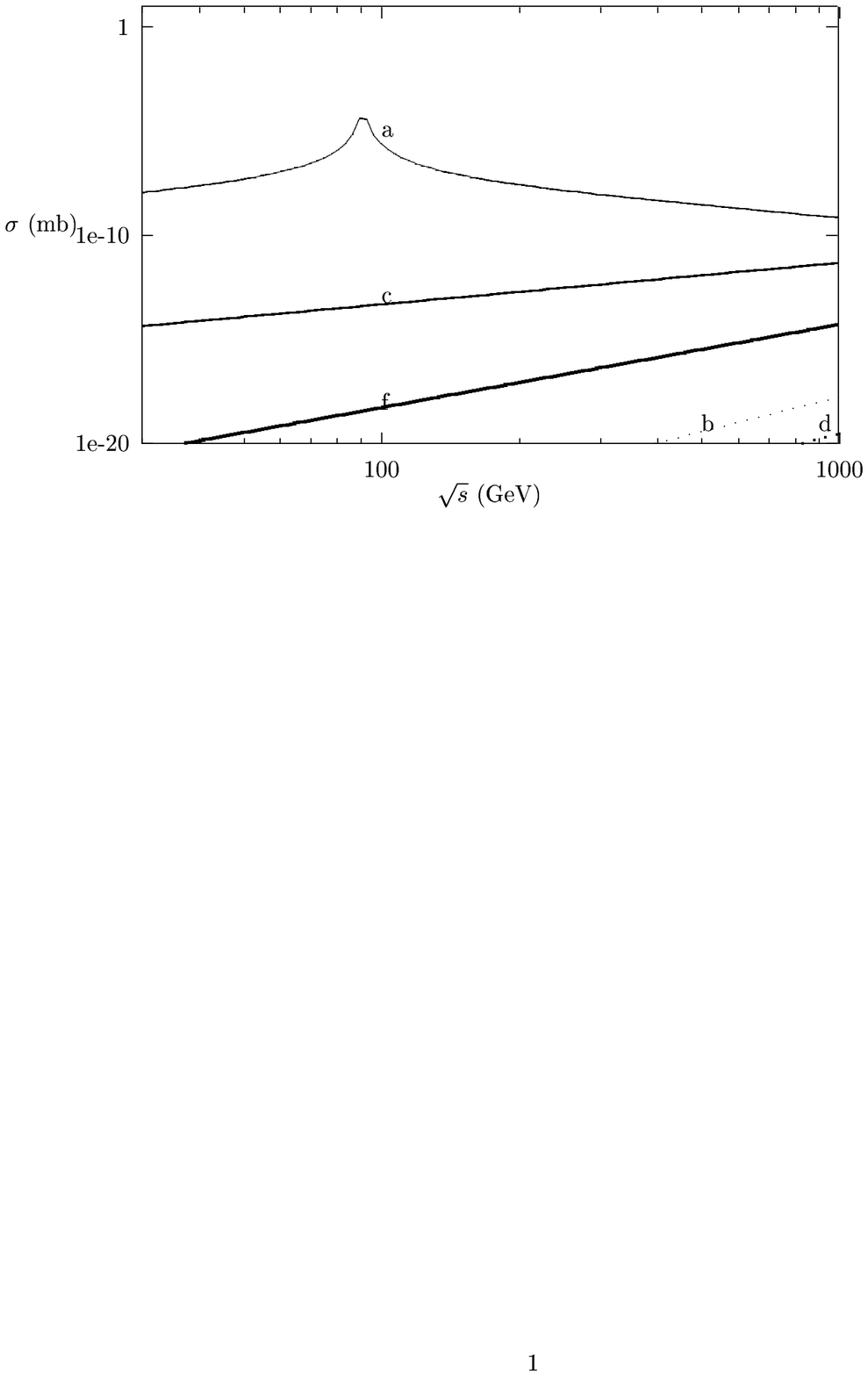}
\caption{Total $\nu\bar{\nu}$ annihilation cross-section as a function of $\sqrt{s}$. The curves are as in Fig.1 and $M_s$ = 30 TeV. }
\end{figure}

\begin{figure}[ht]
\vskip 15truecm
\includegraphics{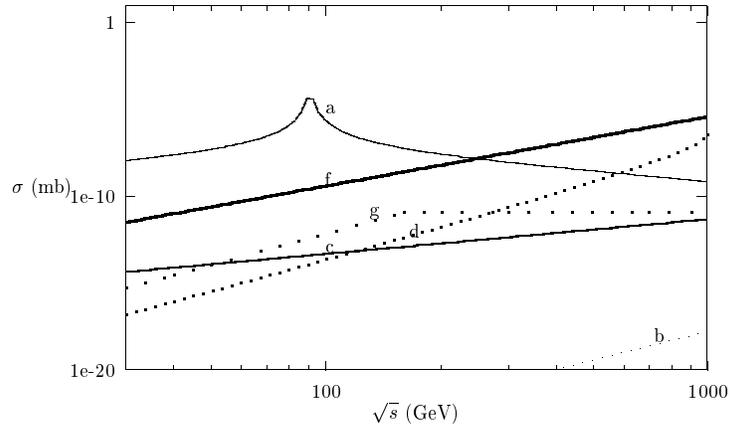}
\caption{Total $\nu\bar{\nu}$ annihilation cross-section as a function of $\sqrt{s}$ for the astrophysically costrained values of the parameters namely, $M_s$ = 30 TeV for n = 2 and $M_s$ = 1 TeV for n =4. Curves a, b, c, d and f are as in Fig.1. Curve g stands for $\sigma^{SM}(\nu\bar{\nu} \longrightarrow \gamma\gamma)$}
\end{figure}

\end{document}